# The Effect of Explainable AI-based Decision Support on Human Task Performance: A Meta-Analysis

**Felix Haag**
University of Bamberg
felix.haag@uni-bamberg.de

**ABSTRACT**

The desirable properties of explanations in information systems have fueled the demands for transparency in artificial intelligence (AI) outputs. To address these demands, the field of explainable AI (XAI) has put forth methods that can support human decision-making by explaining AI outputs. However, current empirical works present inconsistent findings on whether such explanations help to improve users' task performance in decision support systems (DSS). In this paper, we conduct a meta-analysis to explore how XAI affects human performance in classification tasks. Our results show an improvement in task performance through XAI-based decision support, though explanations themselves are not the decisive driver for this improvement. The analysis reveals that the studies' risk of bias moderates the effect of explanations in AI, while the explanation type appears to play only a negligible role. Our findings contribute to the human computer interaction field by enhancing the understanding of human-XAI collaboration in DSS.

**INTRODUCTION**

The advances in artificial intelligence (AI) have changed many aspects of everyday life. This is due to the growing number of tasks where the capabilities of AI are superior to those of humans, fueling fears of full automation (Wilson & Daugherty, 2018). Information systems (IS) scholars have responded to this tension by shifting their focus to the potential that lies in the collaboration between humans and AI. Indeed, research on decision support systems (DSS) has already shown that AI can effectively aid human decision-making (Hemmer et al., 2022). For this collaboration to be adopted, users expect an AI-based DSS to be complementary, adaptive, and transparent in its decisions (Hemmer et al., 2022). In particular, transparency seems to be an essential parameter for success: IS that offer explanations alongside their output have already demonstrated to improve users' decision performance (Gregor & Benbasat, 1999). However, one of the problems of effective but complex AI-based DSS is that they usually rely on sophisticated machine learning (ML) models that are opaque to humans (Meske et al., 2020). To address this issue, research has put forth explainable AI (XAI) methods to explain well-performing, yet complex "black box" AI model decisions (Molnar, 2019).

Employing XAI can be a helpful tool to unleash the full potential of AI in DSS. Indeed, explanations in AI have already proven to be effective in fostering human task performance (Haag et al., 2023). However, while some studies document astounding increases in task performance when decision-making is supported by XAI (e.g., Lai et al., 2020; van der Waa et al., 2021), others find no or even detrimental effects of such decision aid (e.g., Bauer et al., 2023; Carton et al., 2020). Motivated by this ambiguity of results, this paper aims to provide a "snapshot" of the effect of XAI-based decision support on human task performance, taking existing empirical work into account. Our analysis focuses on binary decisions, as most existing studies predominantly employ related classification tasks. Hence, our first Research Question (RQ) asks:

*RQ1: To what extent does XAI-based decision support affect human performance in binary classification tasks?*

When designing DSS, one might question what constitutes a "good" explanation. Recent research in this context suggests that the XAI method employed plays a crucial role for users' task performance (Silva et al., 2023). Our paper therefore attempts to shed light on the effect of popular explanation methods on task performance, namely (i) feature attribution (FA), (ii) counterfactual (CF), and (iii) example-based (EB) explanations (Bauer et al., 2023; van der Waa et al., 2021). Thus, our second RQ therefore asks:

*RQ2: To what extent do (i) FA, (ii) CF, or (iii) EB explanations affect human performance in binary classification tasks?*

To answer our RQs, we conduct a meta-analysis to explore differences in task performance when users are provided with varying types of decision support. In addition, we conduct subgroup analyses to explore whether the risk of bias in studies and the explanation type are decisive moderators for the effect of such support.

**BACKGROUND**

In recent years, XAI research has put forth a wide variety of methods. Given this "XAI jungle" of methods, literature has begun to distinguish between methods (Molnar, 2019). One of these differentiations is the way of how explanations in AI are presented to users (i.e., explanation types). In this context, Herm (2023) adapts user-focused explanation types from Mohseni et al. (2021), that are framed as questions that a user might pose to an ML model to obtain explanations: So-called "*Why?*" explanations help users to understand what input features did or did not lead to a model outcome, while "*Why-not?*" explanations offer contrastive explanations to clarify discrepancies between a prediction and the user's expected outcome. "*How?*" explanations present a holistic representation of the model's inner workings, often visualized through heatmaps and saliency maps (Molnar, 2019). The previously mentioned types are often implemented using *feature-attribution* and *feature importance* (FI) methods, which offer similar interpretations from a user perspective. "*How-to?*" explanations describe minimal hypothetical changes in the input data to shift a prediction towards a desired output. These are typically provided through *counterfactual* methods. Finally, "*What-else?*" explanations present users with similar input instances that lead to the same or similar model result as the original input, usually implemented through *example-based* methods (Herm, 2023; Mohseni et al., 2021).

Building on the interaction with such methods, the field of human-XAI interaction has emerged. Although similar to human–AI collaboration, the focus is on how influencing factors of explanations in AI (e.g., the explanation type) affect resulting outcomes (e.g., task performance) (Schauer, 2024). Despite a growing body of research on human-XAI collaboration in recent years, there is still great uncertainty regarding the effects of such support and the respective explanation types on human task performance.

**METHOD**

The method of our paper is based on the guidelines of vom Brocke et al. (2009) and Higgins and Green (2008).

**Literature Search and Data collection**

Our exploratory string development resulted in the following search string:

*("explainable artificial intelligence" OR "xai" OR "explainable AI" OR "interpretable machine learning" OR "interpretable ml" OR "explainable machine learning" OR "explainable ml" OR ("machine learning" OR "artificial intelligence" OR "AI" AND (interpret\* OR explain\* OR "explanation"))) **AND** ("instance based" OR "example based" OR "counterfactual" OR "hypothetical" OR "causal" OR "anchor" OR "contrastive" OR "feature attribution" OR "feature importance" OR "LIME" OR "SHAP") **AND** ("task performance" OR "decision performance" OR "human accuracy" OR "human performance" OR "user study" OR "empirical study" OR "field experiment" OR "online experiment" OR "human experiment" OR "human evaluation" OR "user evaluation" OR ((behavior\* OR behaviour\*) AND "experiment"))*

To account for a wide range of the available empirical work, we searched three literature databases (EBSCO Host, Web of Science, Scopus) in January 2023. After removing duplicates, we observed a total of 420 papers. Screening titles, abstracts, and keywords for relevance led to 118 articles for full-text assessment. Initially, we filtered full texts by three criteria to include all thematically relevant papers for forward-backward search: articles must involve at least one previously outlined XAI method (C1), include a user study (C2), and report a quantitative human task performance measure (C3). This reduced the pool of relevant articles to 28. From backward and forward citation searches, we identified 37 and 64 additional potentially relevant papers, respectively.

For data collection, we define "AI-based support" as prediction-based assistance and "XAI-based support" as assistance with accompanying explanations. If users receive no support in task fulfillment, we refer to this as "No decision support". To meet the requirements for the following meta-analysis, we further curated the resulting set of papers. For example, we excluded those with within-designs to prevent learning effects and papers without extractable standard deviations for task performance (e.g., from the authors via email or the intersection of studies with Schemmer et al., 2022). Our data collection resulted in 16 studies from 14 papers (Bansal et al., 2021; Wang & Yin, 2021, each containing two studies, respectively).

**Statistical Analysis and Risk of Bias Assessment**

With respect to the target variable of the statistical analyses, we measured task performance as the ratio of binary correct decisions to all decisions. For each comparison between groups, we compute the Standardized Mean Difference (SMD) and also report Hedges' *g* (Hedges, 1981) to avoid a systematic overestimation of the true effect size (i.e., upward bias) for study sample sizes $N \leq 20$ (Higgins & Green, 2008). As the experiments conducted do not share a common true effect size because the experimental setups, tasks, and respective samples considerably vary between studies, we employ random-effects regression models to estimate the mean of effects. In order to avoid repeated comparisons against the same reference group, our analyses divide the sample size of a study's control group by the number of times the group is included in the model (Higgins & Green, 2008).

For quality assessment of the studies considered, our meta-analysis includes an in-depth risk of bias assessment using the current version of the Cochrane tool "RoB 2" (Sterne et al., 2019). The judgement for a study's overall risk of bias results from the answers to signaling questions and is calculated through the RoB2 algorithm. Judgements on the risk of bias can be "Low", "Some concerns" and "High".

**RESULTS**

We start with the comparison of varying decision support (*RQ1*) and then analyze whether the effect of such support is moderated by the explanation type (*RQ2*).

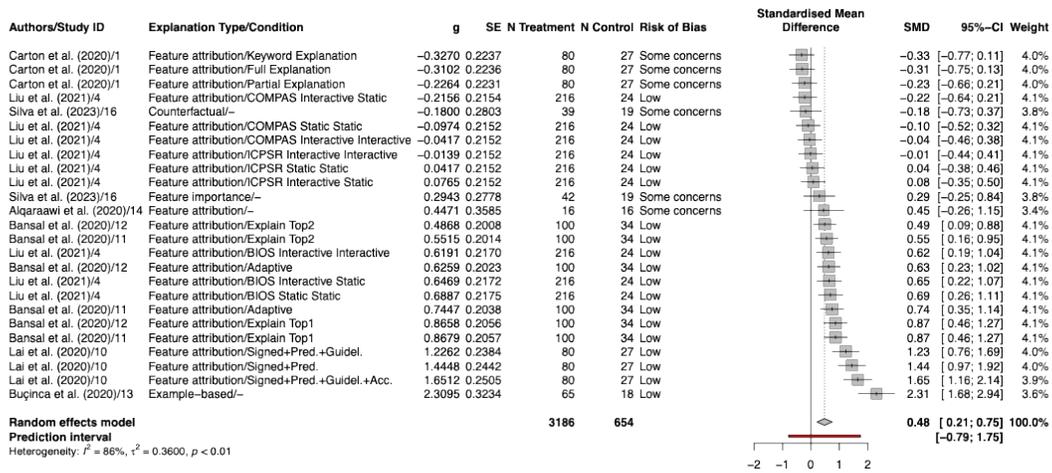

Figure 1. Effect of XAI (Treatment) vs. No (Control) Decision Support

**Effect of XAI-based Decision Support on Human Task Performance**

The analysis for RQ1 begins with the comparison of the main effect between XAI-based decision support and no decision support. Figure 1 displays the forest plot that comprises all conditions in an ascending order according to the study's effect size. We find an increase in the distribution of effects of 0.48 SMD (95% confidence interval (CI) [0.21;0.75]), indicating an effect of "medium size" (Hedges, 1981) through XAI-based decision support. A t-test against the null hypothesis assuming no true effect reveals that the SMD significantly differs from zero ($t$=3.65, $p$=0.001). Hence, considering the body of empirical works included, we find that *XAI-based decision support increases human performance in binary classification tasks, compared to no decision support*.

For the analysis on the comparison between XAI and AI-based decision support (Figure 2), we observe a mean increase in the distribution of effects of 0.09 SMD [-0.01; 0.18] in task performance for XAI-based support when compared to sole AI-based support. This SMD indicates a small enhancing effect (Hedges, 1981). We can reject the null hypothesis that the pooled effect size of XAI-based decision support is zero ($t$=1.86, $p$=0.074) and, therefore, observe a difference in the effect of AI-based and XAI-based decision support but only at the trend level. As the SMD is close to zero and the 95% CI includes zero, we find that *there is no detectable effect of XAI-based decision support on human performance in binary classification tasks, compared to AI-based decision support*.

In addition, we perform a risk of bias assessment using the RoB 2 tool. Deviating from RoB 2, we assume a low risk of bias for studies with no analysis plan. For the case "XAI vs. no decision support", we find that the *effect is not critically influenced by the risk of bias in studies*. However, for the comparison between XAI and AI-based decision support, we observe an increase in task performance for high-risk studies: For studies rated having a "low" risk of bias, XAI-based decision support shows a minor and not statistically significant effect (SMD=0.07, $p$=0.47). Studies assessed with "Some concern" exhibit a slightly negative but also non-significant impact on the effect estimate (SMD=-0.02, $p$=0.89). However, studies with "High" risk of bias yield a considerable and significant increase in task performance (SMD=0.38; $p<0.001$), suggesting an overestimation of the effect of explanations. The test for subgroup differences confirms variations from the studies' risk of bias ($p<0.001$). We therefore conclude that the *studies' risk of bias influences the effect estimate for the case "XAI vs. AI decision support"*.

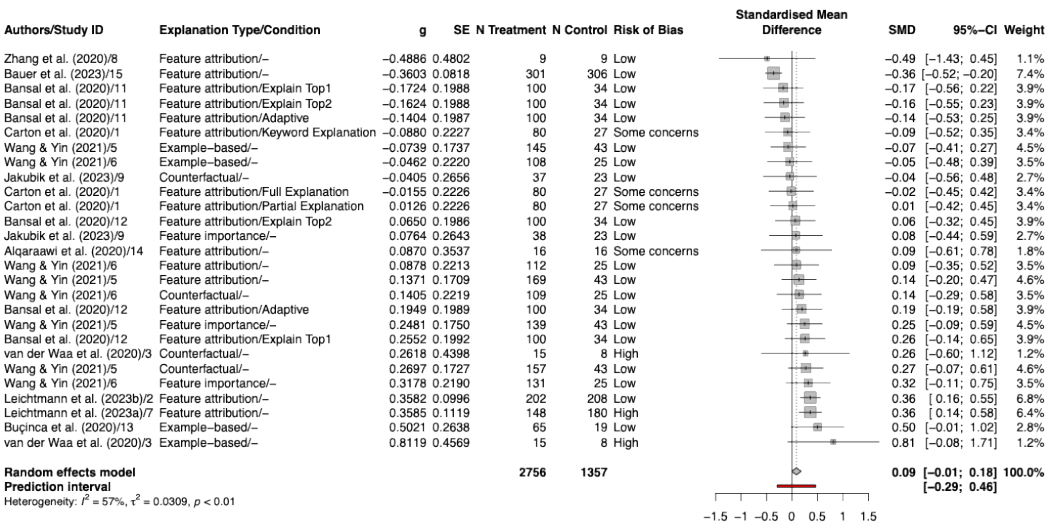

Figure 2. Effect of XAI (Treatment) vs. AI (Control) Decision Support

**Effect of Explanation Types on Human Task Performance**

To address RQ2 of this paper, we focus on the comparison of XAI-based and AI-based support to better understand how the explanation type marginally affects human performance. We find an increase in task performance for all explanation types, as indicated by the positive values of the reported SMDs (Table 1). Despite these increases, none of them show a significant impact on task performance ($p_{type}$>0.05 for all explanation types). The test on subgroup differences also yields *no significant differences between explanation types in human task performance*.

|  | **CF** *How-to* | **FA/FI** *Why/How* | **EB** *What-else* |
|---|---|---|---|
| SMD | 0.17 | 0.06 | 0.18 |
| 95% CI | [-0.04;0.39] | [-0.05;0.18] | [-0.43;0.80] |
| $p_{type}$ | 0.14 | 0.23 | 0.37 |
| $p_{subgroup}$ | 0.42 | | |

**Table 1. Effect of Explanation Types (XAI vs. AI)**

**DISCUSSION AND FUTURE WORK**

We answer RQ1 by finding that *XAI-based decision support increases human performance in binary classification tasks with a medium effect size* when contrasted to groups that receive no decision support. However, explanations in AI-based support do not appear to be the decisive driver for this increase: The comparison with groups that are already equipped with AI-based decision support shows that *there is no detectable effect of XAI-based decision support on human performance*. This outcome corroborates the findings of an earlier meta-analysis conducted by Schemmer et al. (2022), which incorporated nine papers and identified a small, albeit non-significant, advantage of XAI over AI-only decision support. Our study expands upon these results by incorporating 14 papers and conducting in-depth analysis on potential moderators. Interestingly, our risk of bias analysis reveals that high-risk studies lead to an increase of the effect estimate. This result strengthens the argument that explanations in AI-based decision support do not inherently lead to an increase in task performance. With respect to RQ2, we find that *the explanation type does not influence the effect estimate of human task performance* in our sample. One possible reason for this might be that effective support requires more nuanced tailoring to both the user and case. Therefore, our findings suggest that developing effective XAI-based DSS necessitates a broader focus that extends beyond the explanation type.

Despite our best efforts, our meta-analysis has several limitations. First and foremost, one potential weakness results from the limited number of studies included. We also planned to include further subgroup analyses (e.g., to analyze the moderating effect of AI literacy), but were unable to do so due to the small number of studies that report respective measures. A further limitation concerns the generalizability of our results beyond task performance in binary classification tasks. Future research could address this gap by analyzing studies that measure performance as a continuous outcome (see, e.g., Haag et al., 2023) and might also reveal other moderators (e.g., task complexity) for task performance, promoting a more nuanced understanding of human-XAI collaboration in DSS.

**CONCLUSION**

In this paper, we presented a synthesis of empirical work that explores the impact of XAI-based decision support on human performance in binary classification tasks. Our meta-analysis shows that XAI can increase task performance. However, we find that explanations do not inherently lead to task performance increases. By shedding light on how XAI-based decision support affects human task performance, our work adds to the current discourse on human-XAI collaboration in IS (Schauer, 2024).